\documentclass{article}

\usepackage[english]{babel}
\usepackage{authblk}
\usepackage{sidecap}
\usepackage[letterpaper,top=2cm,bottom=2cm,left=3cm,right=3cm,marginparwidth=1.75cm]{geometry}

\usepackage{amsmath}
\usepackage{graphicx}
\usepackage[colorlinks=true, allcolors=blue]{hyperref}
\usepackage{bm}

\newcommand{\td}{\text{d}}

\begin{document}

\title{Coupled shape and spin evolution of small  near spherical asteroids due to global regolith motion}

\author[1]{
Kumar Gaurav}
\author[2]{ Deepayan Banik}
\author[1,3]{ Ishan Sharma}
\date{}

\affil[1]{Applied Mechanics Laboratory, Department of Mechanical Engineering, IIT Kanpur, India } 
\affil[2]{ Department of Physics, University of Toronto, Canada }
\affil[3]{ Space, Planetary $\&$ Astronomical Sciences and Engineering,
IIT Kanpur, India}
\maketitle

\begin{abstract}
Recent space missions have provided substantial evidence of regolith movement on the surfaces of near Earth asteroids.  To investigate this phenomenon, we present a continuum-based model that describes regolith motion on nearly spherical asteroids. The theoretical framework employs a depth-averaged approach, traditionally used for simulating terrestrial landslides, and is extended to include additional terms that account for spherical geometry, shallow topography and the asteroid’s rotation. The governing equations couple the resurfacing process with the asteroid’s spin evolution through angular momentum conservation. The axisymmetric form of these equations is then employed to study the transition of an initially spherical asteroid into a top-shaped.  
\end{abstract}
\section{Introduction}

Regolith—loose, granular material—is found in abundance on small asteroids \cite{lauretta2019unexpected,jawin2022global,tang2024characterization}. These unconsolidated surface materials can be re-distributed by external forces such as meteorite impacts or changes in the asteroid's rotation caused by solar torques. As a result, asteroid surfaces are dynamic, exhibiting features such as filled craters, landslides, equatorial ridge formations, and grain segregation patterns \cite{ghosh2024segregation,jawin2020global}.

One of the most commonly used methods to model regolith motion is the Discrete Element Method (DEM) \cite{Tang,song2024integrated}. Recently, \cite{song2024integrated} introduced a DEM based approach to simulate regolith behavior across an entire asteroid, demonstrating phenomena such as surface motion, mass ejection into orbit, and subsequent reaccumulation on the surface. While their model incorporated multiple physical processes, it was computationally intensive. An alternative modeling strategy employed by \cite{banik_top,gaurav2021granular,gaurav2025regolith} is based on a continuum approach, originally developed by \cite{savage1989motion} to simulate granular flows in laboratory experiments. In this study, we build upon the framework presented in\cite{banik_top} to simulate landslides on a spherical asteroid with shallow basal topography—a reasonable approximation for many naturally occurring top-shaped asteroids. We adopt the governing equations from \cite{gaurav2025regolith} and reformulate them in spherical coordinates, with further simplifications under the assumption of axisymmetry. Additionally, we derive the governing equations for the asteroid’s angular acceleration resulting from axisymmetric regolith motion. Finally, we demonstrate the application of this model in several representative scenarios.

\section{Governing equations}\label{sec:3}
\begin{SCfigure}
  \centering
  \caption{ The schematic shows the asteroid rotating with an angular velocity $\boldsymbol{\omega}$ aligned along an arbitrary axis. Three different coordinate systems: inertial $\{ \mathcal{P},P,\hat{\boldsymbol{E}}_i\}$, body fixed $\{\mathcal{E}_0,G,\hat{\boldsymbol{e}}_i\}$ and curvilinear $\{ \mathcal{E},R,\hat{\boldsymbol{g}}_i\}$ are shown. $\boldsymbol{r} $ is the position vector of any point on the surface w.r.t to the origin G.  A differential volume element is represented by $dV$. }
  \includegraphics[width=0.5\textwidth]%
    {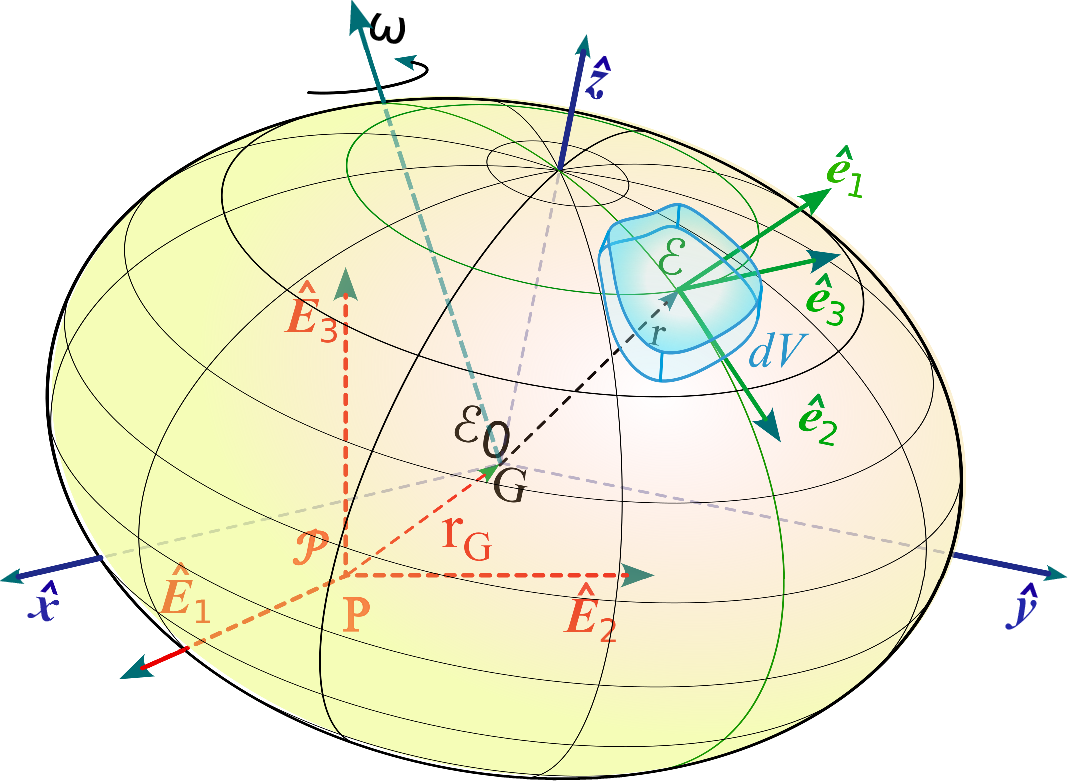}
    \label{fig:oblate}
\end{SCfigure}
The governing equations for flow of regolith on an asteroid rotating about its principal axes are \cite{gaurav2025regolith}
\begin{flalign}
&& &&\nabla\cdot \boldsymbol{u}&=0,\label{eq:cont_2} &&\\
&& && \rho\Dot{\boldsymbol{u}}+\nabla\cdot(\rho \boldsymbol{u}\otimes\boldsymbol{u})&=-\nabla\cdot\boldsymbol{P}+\rho(\boldsymbol{b}-2\boldsymbol{\omega}\times \boldsymbol{u}-\boldsymbol{\alpha}\times\boldsymbol{r})\label{eq:ns_2} &&\\
&&\text{and} &&\boldsymbol{I}\cdot\boldsymbol{\alpha} + \Dot{\boldsymbol{h}} + \Dot{\boldsymbol{I}}\cdot \boldsymbol{\omega}&= - ({\boldsymbol{t}}_s +\Dot{\boldsymbol{I}}_s\cdot \boldsymbol{\omega})  \label{eq:amb_3}, &&
\end{flalign}
where $\boldsymbol{u}$ is the velocity in the coordinate system (CS) $\mathcal{E}$ (see Fig.\ref{fig:oblate}), $\rho$ is the density of the regolith, $\boldsymbol{P}$ is the pressure tensor, $\boldsymbol{\alpha}$ is the angular acceleration of the CB, $\boldsymbol{r}$ is the position vector relative to $G$,   $\boldsymbol{b}=\boldsymbol{b}_0-\boldsymbol{\omega}\times(\boldsymbol{\omega}\times \boldsymbol{r})$ is the total effective gravity with $\boldsymbol{b}_0$ being the gravity of the CB and $\boldsymbol{\omega}\times(\boldsymbol{\omega}\times \boldsymbol{r})$ as the centrifugal acceleration, $2\boldsymbol{\omega}\times \boldsymbol{u}$  is the Coriolis acceleration, $\boldsymbol{I}$ is the moment of inertia of the system about $G$, $\boldsymbol{h}$ is the angular momentum of the regolith in the CS $\mathcal{E}_0$ about  $G$, $\Dot{(\ )}$ represents time derivative in the body fixed CS and  $\Dot{\boldsymbol{I}}_s$ and ${\boldsymbol{t}}_s$  are the rate of loss of moment of inertia and angular momentum due to mass shedding in in the CS $\mathcal{E}$, respectively. 

We also require boundary conditions and a constitutive law to describe flowing regolith. For the former, at the  free top surface of the flow, we will impose the usual kinematic conditions, along with the vanishing of the traction. At the bottom surface, the flow is constrained to adhere to the basal topography, while the traction will be governed by Coulomb's friction law. We note that here we assume no erosion or deposition during flow, which will be incorporated in the future. 

For the constitutive law we employ the {simple rheology}\cite[chap.~2]{sharma2017shapes}\cite{schaeffer1987instability}
\begin{align}
    \boldsymbol{P}=p{\bm 1}-\mu p\frac{\boldsymbol{D}}{|\boldsymbol{D}|},\label{eq:constitutive}
\end{align}
where $p=\text{tr}(\boldsymbol{P})/3$ is the isotropic part of the pressure tensor,  ${\bm 1}$ is the identity tensor, $\boldsymbol{D}$ is the strain rate tensor, $|\boldsymbol{D}|=\sqrt{\text{tr}(\boldsymbol{D}^2)/2}$ and $\mu$ is the dry friction coefficient that characterizes the grains comprising the regolith, and which is taken to be independent of flow properties. We note that \cite{schaeffer1987instability} has demonstrated that the constitutive description is ill-posed in the sense that the perturbation at high wave numbers grows unboundedly (Hadamard unstable). Due to this, the numerical solutions of the system of equations \eqref{eq:cont_2}-\eqref{eq:ns_2} depends on the grid size. At the same time, we show in the App. \ref{sec:stability}, that the depth-averaged equations using the above constitutive relation is well-posed. This happens because only the basal friction appears in the final equations and \eqref{eq:constitutive} is used only for scaling the shear stress in the subsequent analysis. 

\section{Governing equations in spherical coordinates}\label{app:GE}
The continuity equation \eqref{eq:cont_2} and LMB equation \eqref{eq:ns_2} in spherical coordinates become

\begin{align}
    \frac{\partial }{\partial r}\left(u_r r^2 \sin\theta\right)+\frac{\partial }{\partial \theta}\left(u_\theta r \sin\theta\right)+\frac{\partial }{\partial \phi}\left(u_\phi r\right)=0 \label{app:cont_1}\\
   \rho\bigg\{ \frac{\partial}{\partial t}\left(u_r r^2\sin\theta\right)+  \frac{\partial}{\partial r}\left(u_r^2r^2\sin\theta\right)+ \frac{\partial}{\partial \theta}\left(u_r u_\theta r\sin\theta\right)+ \frac{\partial}{\partial \phi}\left(u_r u_\phi r\right)-(u_\theta ^2+u_\phi ^2)r\sin\theta\bigg \}\nonumber\\
   =- \bigg\{ \frac{\partial}{\partial r}\left(P_{rr} r^2\sin\theta\right)+ \frac{\partial}{\partial \theta}\left(P_{r\theta} r\sin\theta\right)+ \frac{\partial}{\partial \phi}\left(P_{r\phi} r\right)-(P_{\theta\theta} +P_{\phi\phi} )r\sin\theta\bigg \}+\rho BF_r r^2\sin\theta\label{app:lmb_1}\\
   \rho\bigg\{ \frac{\partial}{\partial t}\left(u_\theta r^3\sin\theta\right)+  \frac{\partial}{\partial r}\left(u_\theta u_r r^3\sin\theta\right)+ \frac{\partial}{\partial \theta}\left( u_\theta^2 r^2\sin\theta\right)+ \frac{\partial}{\partial \phi}\left(u_\theta u_\phi r^2\right)-u_\phi ^2 r^2 \cos\theta\bigg \}\nonumber\\
   =- \bigg\{ \frac{\partial}{\partial r}\left(P_{r\theta} r^3\sin\theta\right)+ \frac{\partial}{\partial \theta}\left(P_{\theta\theta} r^2\sin\theta\right)+ \frac{\partial}{\partial \phi}\left(P_{\theta\phi} r^2\right)-P_{\phi\phi} r^2\cos\theta\bigg \}+\rho BF_\theta r^3\sin\theta\\
    \rho\bigg\{ \frac{\partial}{\partial t}\left(u_\phi r^3\sin\theta^2\right)+  \frac{\partial}{\partial r}\left(u_\phi u_r r^3\sin\theta^2\right)+ \frac{\partial}{\partial \theta}\left( u_\phi u_\theta r^2\sin\theta^2\right)+ \frac{\partial}{\partial \phi}\left( u_\phi^2 r^2\sin\theta\right)\bigg \}\nonumber\\
   =- \bigg\{ \frac{\partial}{\partial r}\left(P_{r\phi} r^3\sin\theta^2\right)+ \frac{\partial}{\partial \theta}\left(P_{\theta\phi} r^2\sin\theta^2\right)+ \frac{\partial}{\partial \phi}\left(P_{\phi\phi} r^2\sin\theta\right)\bigg \}+\rho BF_\phi r^3\sin\theta^2,\label{app:lmb_3}
\end{align}

where $BF_i$ are the body force terms,
\begin{align}
    BF_r&=b_r+\omega_3^2 r \sin^2\theta+2u_\phi\omega_3\sin\theta,\\
    BF_\theta&=b_\theta+\omega_3^2 r \sin\theta\cos\theta+2u_\phi\omega_3\cos\theta,\\
    BF_\phi&=b_\phi-2\omega_3(u_r\sin\theta+u_\theta\cos\theta)-\alpha_3 r\sin\theta
\end{align}

\subsection{Boundary conditions}\label{app:BC}
The kinematic boundary conditions at top and bottom are 
\begin{align}
    \frac{\partial }{\partial t}F^i(r,\theta,\phi,t) +\boldsymbol{u}^i\cdot \nabla F^i(r,\theta,\phi,t)=0, \hspace{1cm} \hspace{2cm} i=b,s\,\label{eq:kbc}
\end{align}
where superscripts $s$ and $b$ represents top and bottom surfaces, respectively. $F^i(r,\theta,\phi,t)=r-h^i(\theta,\phi,t)=0$ for $i=s,b$ are the equations of the top and bottom surfaces. $h^s(\theta,\phi,t)$ and $h^b(\theta,\phi,t)$ are the height of the top and bottom surfaces. The kinetic boundary conditions at the top is
\begin{align}
    \boldsymbol{P}^s\cdot \boldsymbol{n}^s=0,\label{eq:dbc_1}
\end{align}
where $\boldsymbol{n}^s$ is normal to the top surface and $\boldsymbol{P}^s$ is the pressure tensor at the surface. The kinetic boundary condition at the base is
\begin{align}
    \boldsymbol{P}^b\cdot\boldsymbol{n}^b=(\boldsymbol{n}^b\cdot \boldsymbol{P}^b \cdot \boldsymbol{n}^b)( -\boldsymbol{u}^b/{|\boldsymbol{u}^b|}\mu +\boldsymbol{n}^b), \label{eq:dbc_2}
\end{align}
where $\boldsymbol{u}^b$ and $\boldsymbol{P}^b$ are the velocity and the pressure tensor at the base, respectively. $\mu$ is the friction coefficient at the bottom. We assume $\tan^{-1}(\mu)$ equal to the friction angle of the grains $\delta$.

\section{Non-dimensionalization} 
To facilitate identification of small terms, we non-dimensionalize the governing equations employing relevant physical scales. In the process, we define the two dimensionless parameters
   \begin{align}
    \varepsilon=\frac{H}{R},\hspace{1cm} \text{and} \hspace{1cm} \gamma=\frac{H^b}{R},\label{eq:scales}
   \end{align}
   where $R$ is the {mean radius of the CB}, $H$ is the flow thickness, and $H^b$ is the {topography thickness which is defined as the height of the basal surface above the mean sphere of the CB}. { We assume $\varepsilon<\gamma\ll 1$}.  This means that the asteroid's topography has variations that are small compared to the asteroid's overall size, but significant in the context of individual landslides.  This is assumed to consider the fact that the basal topography is formed due to multiple landslides and hence its thickness grows after each landslide. This is a crucial distinction from the standard avalanche models of \cite{gray1999gravity, banik_top}, where flow and topography were considered of the same order. Introducing $\gamma$ allows us to preserve important curvature terms in our model that, in turn, have implications on the reshaping of the asteroid.
   
As discussed above, the shallow nature of the flow introduces  minimal changes in the body's angular velocity $\omega$. Introducing the perturbation $\tilde{\bm \omega}$, with ${\tilde \omega}/\omega_0 = O(\varepsilon)$, where ${\bm \omega}_0$ is the initial angular velocity of the CB, we  then write
\begin{align}
   {\bm \omega }&= {\bm \omega}_0 + \tilde{\bm \omega} \hspace{1cm} \text{and} \hspace{1cm} 
   {\bm \alpha} \,= \frac{\td {\bm \omega}}{\td t} = \frac{\td \tilde{\bm \omega}}{\td t}. \label{eq:omega}
\end{align} 

 Consequently, the angular acceleration is of order $O(\varepsilon)$, and this simplifies the LMB equations, leading to one-way coupling with the AMB, in contrast to the fully coupled approach of \cite{banik_top}.

 We can rewrite the radial location of any point as, 
\begin{equation}
  r=R+\Tilde{r}  \hspace{1cm }\text{and}\hspace{1cm} h^i(\theta,\phi,t)=R+\tilde{h}^i(\theta,\phi,t),\label{eq:rescale}
\end{equation}
 where $\Tilde{r}$ is the radial distance above the  reference surface
 The thickness of the flow at any point $(\theta,\phi)$ is denoted by $h(\theta,\phi,t)$, s.t.

\begin{align}
    h(\theta,\phi,t)=h^s(\theta,\phi,t)-h^b(\theta,\phi)=\tilde{h}^s(\theta,\phi,t)-\tilde{h}^b(\theta,\phi)\hspace{1cm}\text{and}\hspace{1cm} \frac{h}{R}=O(\varepsilon).
\end{align}

Since we are not considering erosion and deposition during the flow, the bottom surface $h^b$ is independent of time $t$. Following the standard procedures in avalanche dynamics, we non-dimensionalize the equations using the following scales
 \begin{equation}
\begin{array}{cc}  \left\{\Tilde{r},\Tilde{h}^i,h,t,\Tilde{\omega}\right\}=\left\{\gamma R\hat{r},\gamma R \hat{h}^i,\epsilon R \hat{h},\left(R/b\right)^{1/2}\hat{t},\varepsilon\omega_0\hat{\omega}\right\};\\ \hspace{1cm}
\left\{u_r,u_\theta,u_\phi\right\}=(bR)^{1/2}\left\{\varepsilon\hat{u}_r,\hat{u}_\theta,\hat{u}_\phi\right\}; \hspace{2mm}   \left\{\omega_0,\alpha\right\}=\sqrt{\displaystyle\frac{b}{R}}\left\{\hat{\omega}_0,\varepsilon\sqrt{\displaystyle\frac{b}{R}}\hat{\alpha}\right\};\\ \hspace{3mm}
\text{and} \hspace{3mm} \left\{P_{rr},P_{\theta\theta},P_{\phi\phi},P_{r\theta},P_{r\phi},P_{\theta\phi}\right\}=\varepsilon\rho g R\left\{\hat{P}_{rr},\hat{P}_{\theta\theta},\hat{P}_{\phi\phi},\mu\hat{P}_{r\theta},\mu\hat{P}_{r\phi},\mu\hat{P}_{\theta\phi}\right\},
\end{array}\label{eq:non_dim}
\end{equation}
where $b$ is the magnitude of the gravitational acceleration on the surface of sphere, $\hat{()}$ represents the non-dimensional quantity, $\varepsilon$ and $\gamma$ are defined in \eqref{eq:scales} and $\tilde\omega$ is defined in \eqref{eq:omega}. The $\,\hat{}\,$ is dropped in subsequent equations to simplify the notations.
 
\subsection{Governing equations}\label{app:ND_GE}
 Using \eqref{eq:scales}-\eqref{eq:omega},\eqref{eq:rescale}-\eqref{eq:non_dim}, the non-dimensional form of the governing equations \eqref{app:cont_1}-\eqref{app:lmb_3} become
\begin{align}
&\frac{\varepsilon}{\gamma}{\frac {\partial }{\partial r}} \left(   u_r \lambda(\theta,\phi) ^{2}\sin\theta  \right)  + {\frac {\partial }{\partial \theta}} \left( u_\theta   \lambda(\theta,\phi) \sin\theta  \right) +  {\frac {\partial }{\partial \phi}} \left(   u_\phi   \lambda(\theta,\phi)  \right)  =0 \label{eq:ndcont}\\
    &\epsilon \left\{ {\frac {
\partial }{\partial t}} \left( u_r  \lambda(\theta,\phi) ^{2}\sin\theta  \right)  + \frac{\varepsilon}{\gamma} {\frac {\partial }{\partial r}} \left(   u_r^{2} \lambda(\theta,\phi) ^{2}\sin\theta  \right)  + {\frac {\partial }{\partial \theta}} \left( u_r  u_\theta   \lambda(\theta,\phi) \sin\theta  \right) +  {\frac {\partial }{\partial \phi}} \left( u_r  u_\phi   \lambda(\theta,\phi)  \right)\right\}  \nonumber \\
    &\hspace{1cm}= - \frac{\varepsilon}{\gamma}{\frac {\partial }{\partial r}} \left(  P_{rr} \lambda(\theta,\phi) ^{2}\sin\theta  \right) - \mu\,\epsilon\left\{ { \frac {\partial }{\partial \theta}} \left(   P_{r\theta}  \lambda(\theta,\phi) \sin \left( \theta \right)  \right) - {\frac {\partial }{\partial \phi}} \left(   P_{r\phi}   \lambda(\theta,\phi)  \right)  \right\}\nonumber\\ &\hspace{2cm} +\left\{ u_\theta^2 + u_\phi^2 +\varepsilon\left( P_{\theta\theta}+P_{\phi\phi}\right)+BF_r\lambda(\theta,\phi)\right\}\lambda(\theta,\phi)\sin  \theta \label{eq:ndmom1} \\
 &{\frac {\partial }{\partial t}} \left( u_\theta  \lambda(\theta,\phi) ^{3}\sin\theta  \right)  +\frac{\varepsilon}{\gamma}{\frac {\partial }{\partial r}} \left(  u_\theta u_r \lambda(\theta,\phi) ^{3}\sin\theta  \right)  + {\frac {\partial }{\partial \theta}} \left( u_\theta^2   \lambda(\theta,\phi)^2 \sin\theta  \right) +  {\frac {\partial }{\partial \phi}} \left( u_\theta  u_\phi   \lambda(\theta,\phi)^2  \right)  \nonumber \\
    &\hspace{1cm}= -\mu\frac{\varepsilon}{\gamma}{\frac {\partial }{\partial r}} \left(   P_{r\theta} \lambda(\theta,\phi) ^{3}\sin\theta  \right) - \epsilon\left\{ { \frac {\partial }{\partial \theta}} \left(  P_{\theta\theta}  \lambda(\theta,\phi)^2 \sin \left( \theta \right)  \right) + \mu{\frac {\partial }{\partial \phi}} \left(  P_{\theta\phi}   \lambda(\theta,\phi)^2  \right)  \right\}\nonumber\\ &\hspace{2cm} +\left\{  u_\phi^2  +\varepsilon P_{\phi\phi}+BF_\theta\lambda(\theta,\phi)\right\}\lambda(\theta,\phi)^2\cos  \theta   \label{eq:ndmom2}\\
&{\frac {\partial }{\partial t}} \left( u_\phi  \lambda(\theta,\phi) ^{3}\sin^2\theta  \right)  +\frac{\varepsilon}{\gamma}{\frac {\partial }{\partial r}} \left(  u_\phi u_r \lambda(\theta,\phi) ^{3}\sin^2\theta  \right)  + {\frac {\partial }{\partial \theta}} \left( u_\phi u_\theta   \lambda(\theta,\phi)^2 \sin^2\theta  \right)  \nonumber \\ &\hspace{1cm}+  {\frac {\partial }{\partial \phi}} \left(  u_\phi^2   \lambda(\theta,\phi)^2\sin\theta  \right) 
    = -\mu\frac{\varepsilon}{\gamma}{\frac {\partial }{\partial r}}   \left( P_{r\phi}  \lambda(\theta,\phi) ^{3}\sin^2\theta  \right) +  BF_\phi\lambda(\theta,\phi)^3\sin^2 \theta \nonumber\\ &\hspace{2cm}  - \epsilon\left\{ \mu{ \frac {\partial }{\partial \theta}} \left(  P_{\theta\phi}  \lambda(\theta,\phi)^2 \sin^2  \theta   \right) + {\frac {\partial }{\partial \phi}}  \left( P_{\phi\phi}  \lambda(\theta,\phi)^2\sin\theta  \right)  \right\} ,\label{eq:ndmom3}
\end{align}
where $\hat{()}$ is removed for the simplicity of the notation. We have approximated
\begin{align}
   1+\gamma r \approx 1+\gamma h^b(\theta,\phi)=\lambda(\theta,\phi)
\end{align}
and non-dimensional $BF_i$ are,
\begin{align}
    BF_r&=b_r+\omega_0^2\sin^2\theta(1+\varepsilon\omega)^2(\lambda(\theta,\phi))+2u_\phi\omega_0\sin\theta(1+\varepsilon\omega),\label{eq:BF1}\\
    BF_\theta&=b_\theta+\omega_0^2 \sin\theta\cos\theta(1+\varepsilon\omega)^2(\lambda(\theta,\phi)) +2u_\phi\omega_0\cos\theta(1+\varepsilon\omega),\label{eq:BF2}\\
    BF_\phi&=b_\phi-2\omega_0(\varepsilon u_r\sin\theta+u_\theta\cos\theta)(1+\varepsilon\omega)-\varepsilon\alpha_3 \sin\theta(\lambda(\theta,\phi))\label{eq:BF3}.
\end{align}

\subsection{Boundary conditions} \label{app:ND_BC}
The kinematic condition in the spherical coordinates is given by
\begin{align}
   \sin\theta\left( h^i u_r^i-u^i_\theta\frac{\partial h^i}{\partial \theta}\right)-u^i_\phi\frac{\partial h^i}{\partial \phi}=\frac{\partial}{\partial t}\left ( \frac{\left(h^i\right)^2\sin\theta}{2}\right)\hspace{2cm} i=b,s,\label{app:bc_1}
\end{align}
which on non-dimensionalisation yields
 \begin{align}
     \sin\theta\left( \frac{\varepsilon}{\gamma} u_r^i(1+\gamma h^i)- u^i_\theta\frac{\partial h^i}{\partial \theta}\right)-  u^i_\phi\frac{\partial h^i}{\partial \phi}=(1+\gamma h^i)\frac{\partial}{\partial t}\left ( h^i\sin\theta\right)\hspace{2cm} i=b,s. \label{eq:nkbc}
 \end{align}
 
The normal to the surface in spherical coordinates is given by
\begin{align}
    \boldsymbol{n}^i&=\displaystyle\frac{\nabla F^i}{|\nabla F^i|}=\frac{ \left\{\hat{\boldsymbol{g}}_r-\frac{1}{h^i}\frac{\partial h^i}{\partial \theta}\hat{\boldsymbol{g}}_\theta-\frac{1}{h^i\sin\theta}\frac{\partial h^i}{\partial \phi} \hat{\boldsymbol{g}}_\phi\right\}}{\sqrt{\left\{1+\frac{1}{(h^i)^2}\left(\frac{\partial h^i}{\partial \theta}\right)^2+\frac{1}{\left(h^i\sin\theta\right)^2}\left(\frac{\partial h^i}{\partial \phi}\right)^2 \right\}}},\hspace{2cm} i=b,s\label{app:n}.
\end{align}

 Using \eqref{app:n}, we obtain the non-dimensionalised dynamic boundary conditions at the top surface as
 \begin{align}
     \mu\gamma\left(P^s_{r\theta}\sin\theta \frac{\partial h^s}{\partial \theta}+P^s_{r\phi}\frac{\partial h^s}{\partial \phi}\right)-P^s_{rr}\sin\theta(1+\gamma h^s)=0 \label{eq:dbc1_1}\\
     \gamma\left(P^s_{\theta\theta}\sin\theta \frac{\partial h^s}{\partial \theta}+\mu P^s_{\theta\phi}\frac{\partial h^s}{\partial \phi}\right)+\mu P^s_{r\theta}\sin\theta(1+\gamma h^s)=0 \label{eq:dbc1_2}\\
     \gamma\left(\mu P^s_{\theta\phi}\sin\theta \frac{\partial h^s}{\partial \theta}+P^s_{\phi\phi}\frac{\partial h^s}{\partial \phi}\right)-\mu P^s_{r\phi}\sin\theta(1+\gamma h^s)=0 \label{eq:dbc1_3}\\
 \end{align}
 and at the bottom surface as
  \begin{align}
     -\mu\gamma\left(P^b_{r\theta}\sin\theta \frac{\partial h^b}{\partial \theta}+P^b_{r\phi}\frac{\partial h^b}{\partial \phi}\right)+P^b_{rr}\sin\theta(1+\gamma h^b)=\left(\mu Fr_r+P^b_{nn} n^b_r\right)\sin\theta K(\theta,\phi) \label{eq:dbc2_1}\\
     -\gamma\left(P^b_{\theta\theta}\sin\theta \frac{\partial h^b}{\partial \theta}+\mu P_{\theta\phi}\frac{\partial h^b}{\partial \phi}\right)+\mu P^b_{r\theta}\sin\theta(1+\gamma h^b)=\left(\mu Fr_\theta+P^b_{nn}n^b_\theta\right)\sin\theta K(\theta,\phi) \label{eq:dbc2_2}\\
     -\gamma\left(\mu P^b_{\theta\phi}\sin\theta \frac{\partial h^b}{\partial \theta}+P^b_{\phi\phi}\frac{\partial h^b}{\partial \phi}\right)+\mu P^b_{r\phi}\sin\theta(1+\gamma h^b)=\left(\mu Fr_\phi+P^b_{nn}n^b_\phi\right)\sin\theta K(\theta,\phi) \label{eq:dbc2_3}
 \end{align}
where $P^b_{nn}=\boldsymbol{n}^b\cdot \boldsymbol{P}^b \cdot \boldsymbol{n}^b$ is the normal pressure at the base,  $n^b_i$ are components of normal vector in the $i-$ direction and 
\begin{align}
    K(\theta,\phi)=\sqrt{(\gamma h^{b})^2+\left(\gamma\frac{\partial h^{b}}{{\partial}\theta}\right)^{2}+\left(\frac{\gamma}{\sin\theta}\frac{\partial h^b}{\partial\phi}\right)^{2}+2 \gamma  h^{b} + 1}.
\end{align}
$Fr_i$ and $P^b_{nn}n^b_i$ upto first order in $\gamma$ are
\begin{align}
    Fr_r&=-\gamma\,\frac{P^b_{rr}}{\sqrt{(u^b_\theta)^2+(u^b_\phi)^2}\sin\theta}\left(u^b_\theta\sin\theta\frac{\partial h^b}{\partial \theta}+u^b_\phi\frac{\partial h^b}{\partial \phi} \right)\label{eq:Fr1}\\
    Fr_\theta&=\frac{u^b_{\theta}}{\sqrt{(u^b_\theta)^2+(u^b_\phi)^2}\sin\theta}\left\{2\mu\gamma\left( P^b_{r\theta}\sin\theta\frac{\partial h^b}{\partial \theta}+P^b_{r\phi}\frac{\partial h^b}{\partial \phi} \right)-P^b_{rr}\sin\theta\right\}\label{eq:Fr2}\\
     Fr_\phi&=\frac{u^b_{\phi}}{\sqrt{(u^b_\theta)^2+(u^b_\phi)^2}\sin\theta}\left\{2\mu\gamma\left( P^b_{r\theta}\sin\theta\frac{\partial h^b}{\partial \theta}+P^b_{r\phi}\frac{\partial h^b}{\partial \phi} \right)-P^b_{rr}\sin\theta\right\}\label{eq:Fr3}\\
     P^b_{nn} n^b_r&= -\frac{1}{\sin\theta}\left\{2\mu\gamma\left( P^b_{r\theta}\sin\theta\frac{\partial h^b}{\partial \theta}+P^b_{r\phi}\frac{\partial h^b}{\partial \phi} \right)-P^b_{rr}\sin\theta\right\}\\
     P^b_{nn} n^b_\theta&= -\gamma\frac{\partial h^b}{\partial \theta}P^b_{rr}\\
      P^b_{nn} n^b_\phi&= -\frac{\gamma}{\sin\theta}\frac{\partial h^b}{\partial \phi}P^b_{rr}\label{eq:Pb3}
\end{align}

To close the system of equations, we use a simple constitutive relation \eqref{eq:constitutive} for the regolith. Constitutive relation \eqref{eq:constitutive} can be non-dimensionalized, and terms of $O(\varepsilon)$ and $O(\gamma)$ can be neglected since the components of pressure tensor are multiplied by $\varepsilon$ and $\gamma$ everywhere in the equations. The non-dimensionalized form of constitutive law at leading order becomes
\begin{align}
    P_{rr}=P_{\theta\theta}=P_{\phi\phi}=p+O(\varepsilon,\gamma),\hspace{1cm} P_{r\theta}= -p\frac{\displaystyle\frac{\partial u_\theta}{\partial r}}{\sqrt{\left(\displaystyle\frac{\partial u_\theta}{\partial r}\right)^2+\left(\displaystyle\frac{\partial u_\phi}{\partial r}\right)^2}}+O(\varepsilon,\gamma),\label{eq:P1}\\
    P_{\theta\phi}=O(\varepsilon,\gamma) \hspace{1cm}\text{and}\hspace{1cm} P_{r\phi}= -p\frac{\displaystyle\frac{\partial u_\phi}{\partial r}}{\sqrt{\left(\displaystyle\frac{\partial u_\theta}{\partial r}\right)^2+\left(\displaystyle\frac{\partial u_\phi}{\partial r}\right)^2}}+O(\varepsilon,\gamma).\label{eq:P2}
\end{align}

\section{Depth averaging} 
The shallowness of the flow allows us to integrate the governing equation through the depth, i.e. along the  $r-$ direction \cite{savage1989motion, gray1999gravity, gaurav2021granular}. We thus define the depth-averaged quantities
\begin{equation}
   \overline{v}(\theta,\phi,t) = \frac{1}{h}\int^{h^s}_{h^b} v(r,\theta,\phi,t) \, \td r \ \  \text{and} \ \ 
   \overline{vw} = \beta_{vw}(\theta,\phi,t) \overline{v} \,\overline{w} \label{eq:vw}
\end{equation}
where $\bar{(\cdot)}$ identifies depth-averaged quantities, $z= h^s$ and $z=h^b$ locate the top surface and basal topography, respectively,  $h$ is the flow height and the  depth-variation parameter $\beta_{vw}$ is introduced to express the average of the product of velocites as the product of their individual averages. For granular flows, we often approximate $\beta_{vw}$ as unity \cite{gray1999gravity}, which assumes a plug-like flow - {a fairly good assumption}\cite{gdr2004dense} -- thereby allowing us to equate  basal and depth-averaged velocities.

The governing equations \eqref{eq:ndcont}-\eqref{eq:ndmom3} can be depth averaged using \eqref{eq:vw}  with the boundary conditions \eqref{eq:nkbc}-\eqref{eq:dbc2_3}.  The depth averaged momentum equation in $r-$direction at the leading order in $\gamma$  becomes
  \begin{align}
       P_{rr}=-&\left( \frac{\overline{u_\theta^2} + \overline{u_\phi^2}}{\lambda}+\overline{b}_r+\omega_0^2  \sin^2\theta+2\overline{u}_\phi\omega_0\sin\theta\right)\left(1-\mu\gamma\left(\hat{u}_\theta^b\frac{\partial h^b}{\partial \theta}+\frac{\hat{u}^b_\phi}{\sin\theta}\frac{\partial h^b}{\partial \phi}
       \right)\right)h+O(\varepsilon,\gamma^2) \nonumber,
 \end{align}
where $\hat{u}^b_\theta$ and $\hat{u}^b_\phi$ is defined in \eqref{eq:uhat}. Equating the base velocity with the depth averaged velocity and further simplifications  yields \eqref{eq:pr}.

Similarly, depth-averaging the continuity equation \eqref{eq:ndcont}, momentum equations in $\theta-$direction \eqref{eq:ndmom2} and $\phi-$direction \eqref{eq:ndmom3} using \eqref{eq:vw} and boundary conditions \eqref{eq:nkbc}-\eqref{eq:dbc2_3}, we obtain

\begin{align}
    &\frac{\partial }{\partial  t}\left\{h\lambda^2\sin\theta \right\}+\frac{\partial }{\partial \theta}\left\{\overline{u}_\theta h\lambda\sin\theta\right\}+\frac{\partial}{\partial \phi}\left\{\overline{u}_\phi h\lambda\right\}=0\label{eq:vwcont}\\
    &\frac{\partial }{\partial t}\left\{\overline{u}_\theta h\lambda^3\sin\theta \right\}+\frac{\partial }{\partial \theta}\left\{( \overline{u_\theta^2}+\epsilon \overline{P}_{\theta\theta})h\lambda^2\sin\theta\right\}+\frac{\partial }{\partial \phi}\left\{( \overline{u_\theta u_\phi}+\epsilon\mu \overline{P}_{\theta\phi})h\lambda^2\right\}\nonumber\\ 
        &\hspace{2cm}= (\epsilon \overline{P_{\phi\phi}}+\overline{u^2_\phi})h\lambda^2\cos\theta+{\left(\mu Fr_\theta-\gamma\frac{\partial h^b}{\partial \theta}P^b_{rr}\right)\lambda^3\sin\theta}+\overline{BF}_\theta\lambda^3 h \sin\theta\label{eq:vwmom1}\\
    &\frac{\partial }{\partial t}\left\{\overline{u}_\phi h\lambda^3\sin^2\theta \right\}+\frac{\partial }{\partial \theta}\left\{( \overline{u_\phi u_\theta}+\epsilon\mu\overline{P}_{\theta\phi})h\lambda^2\sin^2\theta\right\}+\frac{\partial }{\partial \phi}\left\{( \overline{ u^2_\phi}+\epsilon \overline{P}_{\phi\phi})h\lambda^2\sin\theta\right\}\nonumber\\ 
        &\hspace{2cm}= {\left(\mu Fr_\phi-\frac{\gamma}{\sin\theta}\frac{\partial h^b}{\partial \phi}P^b_{rr}\right)\lambda^3\sin^2\theta}+\overline{BF}_\phi\lambda^3 h \sin^2\theta,\label{eq:vwmom2}
\end{align}
where $O(\varepsilon^2)$ terms are neglected. Furthermore, following the general practice in the avalanche dynamics equations, we will ignore the $O(\varepsilon)$ terms in source terms and keep it in flux terms. 
Depth averaging the equations \eqref{eq:BF2},\eqref{eq:BF3},\eqref{eq:P1} and \eqref{eq:P2} using \eqref{eq:pr} and substituting in the depth averaged momentum equations \eqref{eq:vwmom1} and \eqref{eq:vwmom2} along with \eqref{eq:Fr2} and \eqref{eq:Fr3} , we obtain
\begin{align}
      &\frac{\partial }{\partial t}\left\{\overline{u}_\theta h\lambda^3\sin\theta \right\}+\frac{\partial }{\partial \theta}\left\{\left( \overline{u_\theta^2}+\epsilon \psi \frac{h}{2}\right)h\lambda^2\sin\theta\right\}+\frac{\partial }{\partial \phi}\left\{ \overline{u_\theta u_\phi}h\lambda^2\right\}=  \nonumber\\ & 
     \left\{ -\left(\mu  \hat{u}^b_{\theta}+\gamma\frac{\partial h^b}{\partial \theta}\right)\psi   +\frac{\overline{u^2_\phi}}{\lambda}\cot\theta+\left(b_\theta+2\overline{u}_\phi\omega_0\cos\theta+\omega_0^2 \lambda\sin\theta\cos\theta\right)\right\}\lambda^3 h \sin\theta+O(\varepsilon,\gamma^2,\mu^2\gamma)\nonumber\\
    &\frac{\partial }{\partial t}\left\{\overline{u}_\phi h\lambda^3\sin^2\theta \right\}+\frac{\partial }{\partial \theta}\left\{ \overline{u_\phi u_\theta}h\lambda^2\sin^2\theta\right\}+\frac{\partial }{\partial \phi}\left\{\left( \overline{ u^2_\phi}+\epsilon \psi \frac{h}{2}\right)h\lambda^2\sin\theta\right\}= \nonumber\\   &\hspace{2cm}
    \left\{-\left(\mu  \hat{u}^b_{\phi}+\frac{\gamma}{\sin\theta}\frac{\partial h^b}{\partial \phi}\right)\psi +\left\{b_\phi-2\omega_0\overline{u}_\theta\cos\theta\right\}\right\}\lambda^3 h \sin\theta+O(\varepsilon,\gamma^2,\mu^2\gamma).\nonumber
\end{align}

\section{Axisymmetry } 
Impact-induced landsliding  is not inherently symmetric. Nevertheless, at low rotation rates the regolith flow will be largely symmetric about the impact point. Similarly, for  impacts that are large enough to excite global reverberations and, consequently, surface-spanning regolith motion, which we may expect to be symmetric about the rotation axis, at least to leading order of approximation. Here, we limit ourselves to the axisymmetric landslides. Doing so greatly simplifies the presentation, without diminishing any of the interconnected physics in the present framework. Non-axisymmetric landsliding will be added in as part of the next round of improvements. Previous simulations, e.g. by \cite{tardivel2018equatorial} to study the formation of rocky equators and by \cite{vance2022possible} to investigate top shape formation due to mass shedding, also utilized axisymmetry. Enforcing axisymmetry implies that $\overline{v}(\theta,\phi,t) \approx \overline{v}(\theta,t)$, thereby reducing the spatial dimensionality to one -- the $\theta$-direction. This   simplifies the LMB \eqref{eq:ns_2} significantly.

\noindent Introducing the above simplifications, the $r-$ direction LMB up to order $O(\gamma)$ reduces to 
  \begin{align}
       P_{rr}(h^b)=\left(1-\gamma\mu\hat{\overline{u}}_\theta \frac{\td h^b}{\td \theta}\right) \psi(\theta,t)h + O(\varepsilon,\gamma^2), \label{eq:pr}
   \end{align}
where
   \begin{align}
       \psi(\theta,t)h=- \left(\frac{\overline{u}_\theta^2+ \overline{u}_\phi^2}{\lambda} +\overline{b}_r+\omega_0^2  \sin^2\theta+2\overline{u}_\phi\omega_0\sin\theta\right)h
 \end{align}
 is the effective vertical pressure which consists of the hydrostatic pressure modified by the effects of rotation and the curvature of the CB -- see \cite{banik_top} for further discussion, $\lambda(\theta)=1+\gamma h^b(\theta)$ and

\begin{equation}
 \hspace{1cm} \hat{\overline{u}}_i=\frac{\overline{u}_i}{\sqrt{(\overline{u}_\theta)^2+(\overline{u}_\phi)^2}} \hspace{1cm} i=\theta,\phi. \label{eq:uhat}
\end{equation}
The correction factor multiplying $\psi$ is due to the fact that the normal to the basal topography is not along the radial direction and hence a component of vertical pressure appears along radial direction; cf. \eqref{app:n}.  As \cite{banik_top} did not account for basal topography, this term was not present in their analysis.

Assuming $\beta_{vw}=1$ in \eqref{eq:vw} and assuming $\hat
{u}^b_\theta=\hat{u}_\theta$ and $\hat{u}^b_\theta=\hat{u}_\theta$, we obtain the governing equations  as
\begin{flalign}
     && \frac{\partial }{\partial  t}\left\{h\lambda^2\sin\theta \right\}+\frac{\partial }{\partial \theta}\left\{\overline{u}_\theta h\lambda\sin\theta\right\}=0,&& \label{eq:cont_fin}\\
      &&  \frac{\partial }{\partial t}\left\{\overline{u}_\theta h\lambda^3\sin\theta \right\}+\frac{\partial }{\partial \theta}\left\{\left( \overline{u}_\theta^2+\epsilon \psi \frac{h}{2}\right)h\lambda^2\sin\theta\right\}=  \lambda^3 h \sin\theta \dots && \nonumber\\  
      &&\dots\left\{ -\left(\mu  \hat{\overline{u}}_{\theta}+\gamma\frac{\partial h^b}{\partial \theta}\right)\psi   +\frac{\overline{u}^2_\phi}{\lambda}\cot\theta+\left(b_\theta+2\overline{u}_\phi\omega_0\cos\theta+\omega_0^2 \lambda\sin\theta\cos\theta\right)\right\} && \label{eq:vwmom1_fin}\\
    \text{and}&& \hspace{-1mm}  \frac{\partial }{\partial t}\left\{\overline{u}_\phi h\lambda^3\sin^2\theta \right\}+\frac{\partial }{\partial \theta}\left\{ \overline{u}_\phi \overline{u}_\theta h\lambda^2\sin^2\theta\right\}= 
    -\left\{\mu  \hat{\overline{u}}_\phi\psi +2\omega_0\overline{u}_\theta\cos\theta\right\}\lambda^3 h \sin^2\theta; &&\label{eq:vwmom2_fin}
\end{flalign}
 Note that the above equations do not contain $ \omega$ or $\alpha$ and may be solved independently of the AMB. This represents a minor departure from the approach outlined in \cite{banik_top}. This divergence occurred because we neglected terms of $O(\varepsilon)$, while retaining significant {$O(\gamma)$ terms that arose from variations in the curvature of the CB}.

From \eqref{eq:dbc2_1} we find that to the leading order basal pressure $P_{nn}^b$, i.e. pressure normal to the basal topography,  equals $P_{rr}(h^b)$ given by \eqref{eq:pr}. Thus, we assume that mass shedding commences when \[P_{rr}(h^b) < 0;\] this is the {\em mass shedding criterion}. {It  signifies that the initiation of mass loss  from the body's surface commence when the velocity of grains makes the basal pressure negative \cite{gaurav2021granular}}.

\section{Angular momentum balance}

The angular momentum balance equation \eqref{eq:amb_3} in the spherical coordinates  becomes
\begin{align}
    \displaystyle\int^{2\pi}_0\displaystyle\int^{\pi}_0\Bigg\{\int_0^{R+h^s}\rho r^4\left(\boldsymbol{1}-\hat{\boldsymbol{g}}_r\otimes\hat{\boldsymbol{g}}_r \right)\boldsymbol{\alpha}dr+\frac{d}{dt} \int_0^{R+h^s} r^3\hat{\boldsymbol{g}}_r\times \boldsymbol{u}dr\ \  +\nonumber \\
    \left(\frac{d}{dt} \int_{0}^{R+h^s}\rho r^4\left(\boldsymbol{1}-\hat{\boldsymbol{g}}_r\otimes\hat{\boldsymbol{g}}_r \right)dr\right)\boldsymbol{\omega}\Bigg\} \,\sin{\theta}\td\theta d\phi =- ({\boldsymbol{t}}_s +\Dot{{I}}_s\cdot \boldsymbol{\omega}) , \label{app:AMB:1}
\end{align}
which upon using \eqref{eq:rescale}-\eqref{eq:non_dim} and for the case of axisymmetry simplifies to 
\begin{align}
      &\left\{ \int^{\pi}_0\frac{2\pi}{5} \left[(1+\gamma h^s)^5-1\right]\sin^3\theta \,\td\theta+\frac{8\pi}{15}\right\}\varepsilon\alpha +\frac{\pi}{2}\frac{d}{dt} \int^{\pi}_0 \overline{u}_\phi \left[(1+\gamma h^s)^4-(1+\gamma h^b)^4\right]\sin^2\theta \,\td\theta \nonumber\\ &\hspace{2cm}+ \,\frac{2\pi}{5}\omega_0(1+\varepsilon\omega)\frac{d}{dt} \int_{0}^{\pi} \left[(1+\gamma h^s)^5-1\right]\sin^3\theta  \,\td\theta + \varepsilon({{t}}_s +\Dot{{I}}_s\omega_0(1+\varepsilon\omega)) \approx 0,\nonumber
\end{align}
where $\boldsymbol{t}_s=t_s\hat{\boldsymbol{e}}_3$ is assumed to be order $\varepsilon$. Here, we have used the assumption $\overline{vw}\approx\overline{v}\,\overline{w}$. The above equation can be linearized and using the fact that $\partial h^b/\partial t=0$, we obtain 

\begin{align}
    \alpha&=-\frac{1}{C}\left\{\frac{d}{dt}\int_0^\pi \overline{u}_\phi h\lambda^3\sin^2\theta \,\td\theta+\omega_0\frac{d}{dt}\int_0^\pi h\lambda^4\sin^3\theta \,\td\theta +{t}_s+\dot{I}_s\omega_0\right\},\label{eq:amb_lin}
\end{align}
 where we have neglected $O(\varepsilon)$ terms and
\begin{equation}
    C=\frac{4}{15}\left(1+\frac{15}{4}\int^\pi_0\gamma h^b\sin^3\theta \,\td\theta\right)
\end{equation}
is the moment of inertia of CB about the principal axis.

 The AMB equation \eqref{eq:amb_lin} can be further simplified using continuity and momentum balance equations. The limits of the integration in \eqref{eq:amb_lin} are independent of $t$, and hence the derivative can be taken inside. Now using the continuity equation \eqref{eq:cont_fin} and $\phi-$direction momentum equation \eqref{eq:vwmom2_fin}, we obtain
\begin{align}
    \alpha&=\frac{15}{4}\Bigg\{\int_0^\pi\frac{\partial }{\partial \theta}\left\{ \overline{u_\phi}\overline{ u_\theta}h\lambda^2\sin^2\theta\right\}-
    \left\{-\mu  \hat{\overline{u}}^b\psi +2\omega_0\overline{u}_\theta\cos\theta\right\}\lambda^3 h \sin^2\theta\, \,\td\theta  \nonumber \\
    &+\omega_0\int_0^\pi \frac{\partial }{\partial\theta}(\overline{u}_\theta h \lambda \sin\theta)\lambda^2\sin^2\theta\, \,\td\theta-T_s\Bigg\}\left(1+\frac{3}{4}\int^\pi_0(\lambda^5-1+5\varepsilon\lambda^4 h)\sin^3\theta \,\td\theta\right)^{-1},\nonumber
\end{align}
which upon using the integration by parts and \eqref{eq:cont_fin}-\eqref{eq:vwmom2_fin}, we obtain 

\begin{align}
    \alpha= \frac{1}{C}\left\{\int_0^\pi
    \mu  {\hat{\overline{u}}_\phi}\psi \lambda^3 h \sin^2\theta \,\td\theta-{t}_s-\dot{I}_s\omega_0\right\}.\label{eq:amb_fin}
\end{align}

While deriving \eqref{eq:amb_fin}, we have used the kinematic boundary condition \eqref{eq:nkbc}, which when used for base reduces for the axisymmetric case to a simpler form
\begin{align}
    \gamma u_\theta^b\frac{\partial h^b}{\partial \theta}=\varepsilon u^b_r.
\end{align}

The form of the AMB in \eqref{eq:amb_lin} allows us to determine the angular acceleration $\alpha$ at any instant of time in terms of the average azimuthal velocity $\bar{u}_\phi$ and the landslide's depth $h$. At the same time, the expression \eqref{eq:amb_fin} for the AMB reveals that, in the absence of external torque, the angular momentum of the CB is modified solely by interaction with the flowing regolith -- the integral within \eqref{eq:amb_fin} computes the cumulative torque due to basal friction in the $\phi-$direction. Thus, we could also have obtained \eqref{eq:amb_fin} by equating the total moment induced on the CB by the basal drag of the regolith flow with the rate of change of its angular momentum. 

 To determine the change $\delta\omega$ in the asteroid's angular velocity  due to regolith flow, we integrate  \eqref{eq:amb_lin} over the duration $t_{f}$ of the landslide. This yields
\begin{equation}
    \delta\omega= -\frac{\omega_0}{C}\left(\left[\int_0^\pi h\lambda^4\sin^3\theta \,\td\theta \right]^{t=t_{f}}_{t=0}+\delta\tilde{H}_s\right) \label{eq:AMB_SC},
\end{equation}
where $\delta \tilde{H}_s$ represents the total loss of angular momentum due to mass shedding. We may also obtain \eqref{eq:AMB_SC} directly by equating the initial and final angular momenta of the isolated system comprised of the CB and the flowing regolith.

\section{Results}\label{sec:shape}

\begin{figure}[!ht]
    \centering
    \includegraphics[width=\textwidth]{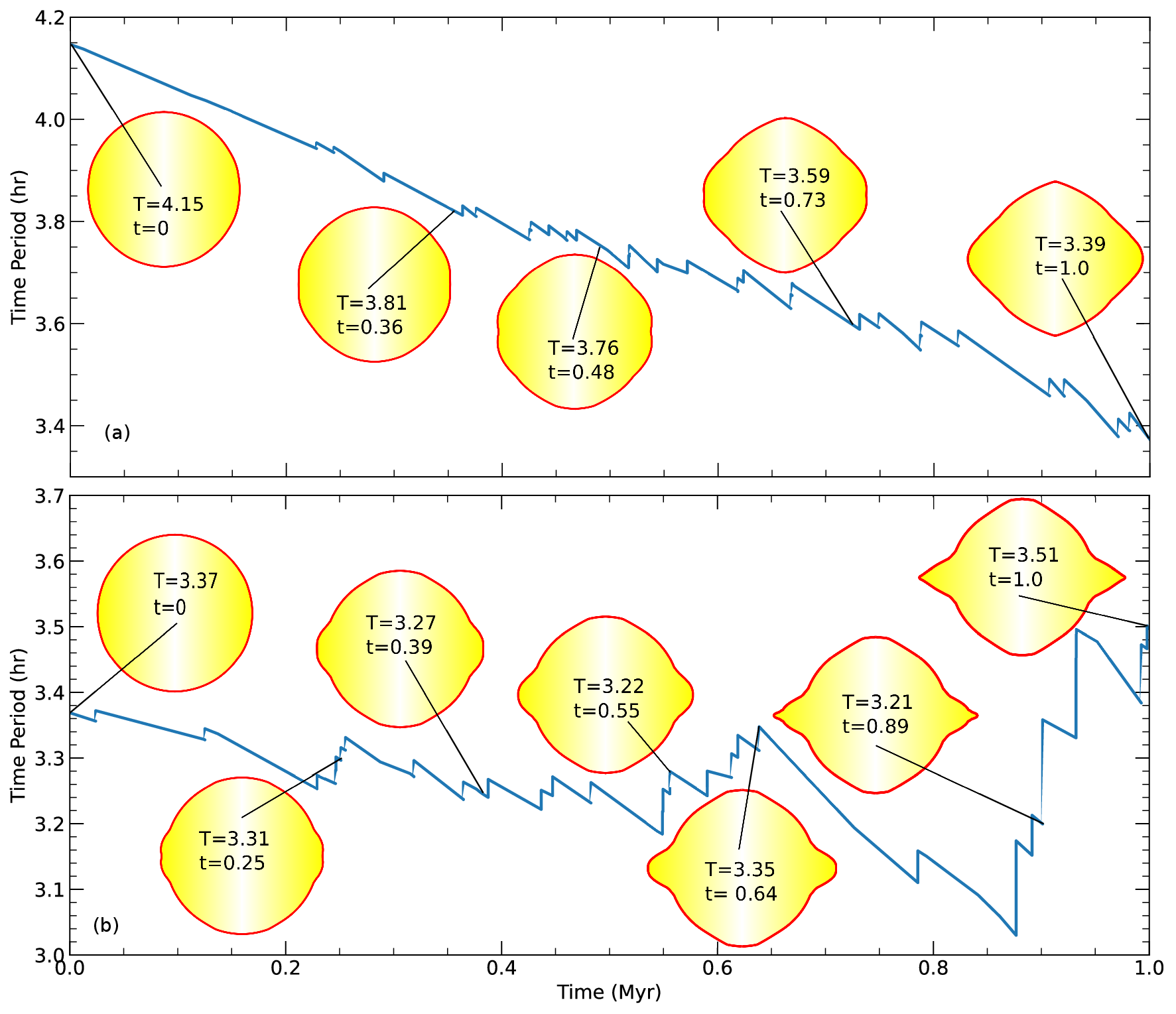}
    \caption{Evolution of shape and spin driven by increasing solar torque and multiple landslides. The evolution time and the rotation period are $t$ and $T$, respectively. (a) Simulation with friction angle $15^\circ$ and smaller initial rotation rate ($T=4.15$ hrs) (b)  Simulation with friction angle $20^\circ$ and faster initial rotation rate ($T=3.37$ hrs).}
    \label{fig:shape1}
\end{figure}

The regolith flow model described above can be integrated into the framework proposed by \cite{gaurav2025regolith}, which accounts for both the asteroid’s collisional history and the effects of solar torque that spins up/down the asteroid. After each impact event, a global-scale landslide is simulated using the regolith flow model. These landslides progressively modify the asteroid's surface, with each subsequent event occurring on the reshaped terrain, ultimately leading to noticeable changes in the asteroid’s overall shape over long timescale of million years (Myrs).

Figure \ref{fig:shape1}a shows how an initially spherical asteroid, spinning with a period of 4.15 hours, evolves over time. The change in the spin shown in the figure is due to combined effects of external solar torque and regolith motion. The smooth part of the spin evolution is due to solar torques and the jump is due to surface motion. As the asteroid spins up, regolith slowly moves towards the equator, creating a "top" shape. This movement is primarily caused by the tangential component of centrifugal force. However, near the equator, this force weakens, delaying the initial formation of a bulge (noticeable at 0.5 Myr in Fig. \ref{fig:shape1}a). As the asteroid's spin increases, the centrifugal force's tangential component grows stronger.  At the same time, increasing oblateness weakens normal gravity, reducing the normal reaction force and consequently, frictional resistance.  This combination of stronger force and weaker friction accelerates bulge formation after t=0.73.

\begin{figure}[!ht]
    \centering
    \includegraphics[width=\textwidth]{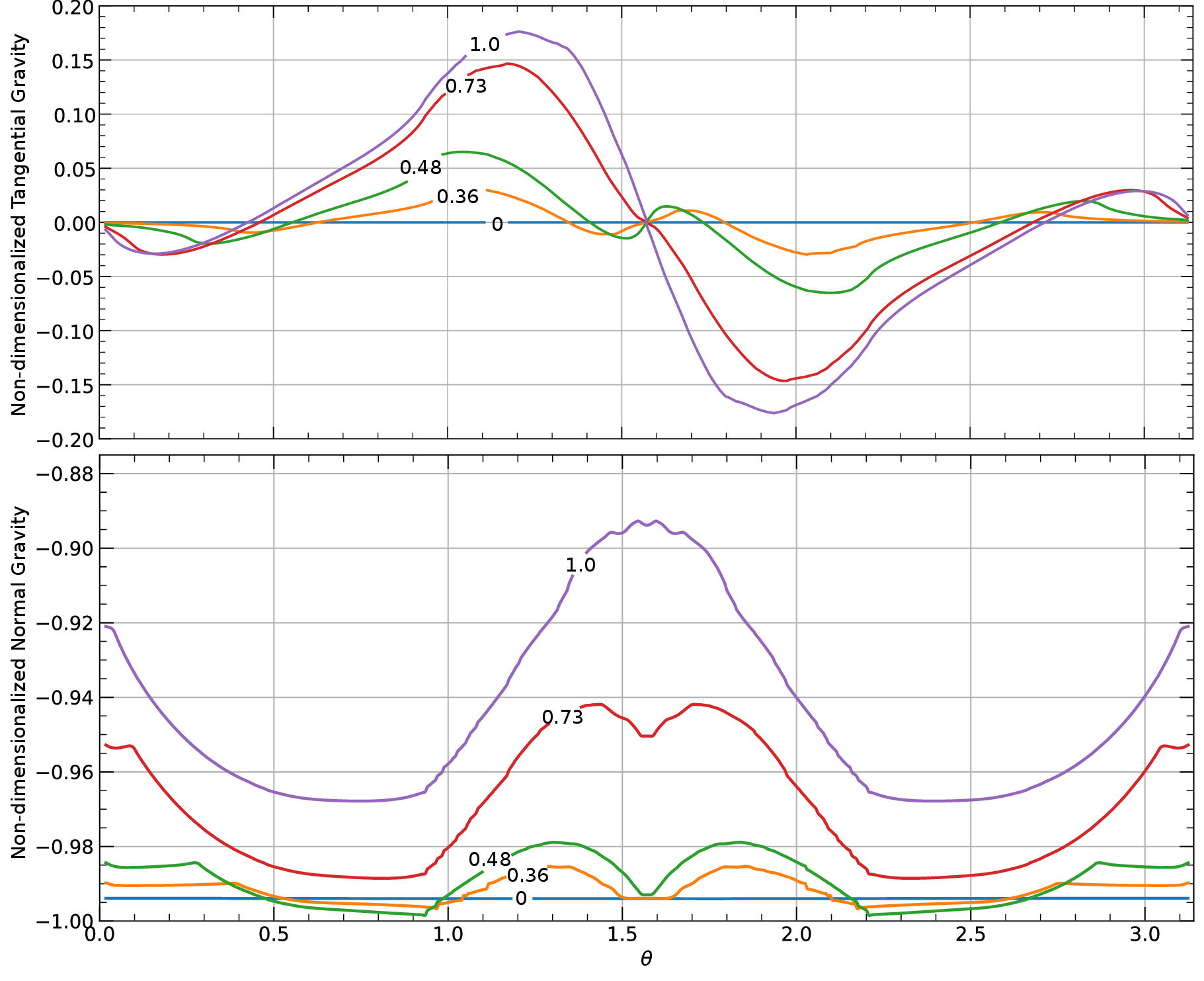}
    \caption{Evolution of tangential (top) and normal (bottom) gravity for the shape evolution shown in the plot \ref{fig:shape1}a. $\theta$ is the latitude and label on each curve represents the time in Myr.}
    \label{fig:grav}
\end{figure}

Changes in an asteroid's shape directly impact its gravitational field, influencing how regolith moves. As seen in Figure \ref{fig:grav}, normal gravity decreases over time while tangential gravity emerges. A perfectly spherical, non-rotating asteroid has equilibrium at every point ($\theta$). As the asteroid's shape evolves, only a specific latitude in each hemisphere remains in equilibrium. Points near the equator maintain stable equilibrium, while those near the poles become unstable (see \ref{fig:grav}a). Regolith accumulates near these stable equatorial points, initially delaying the formation of a bulge. As the asteroid becomes more oblate, the equatorial equilibrium points vanish, and the increased tangential gravity accelerates bulge formation by pulling material towards the equator.
 
 The evolution of the gravity field presents a stark contrast with the findings of \cite{banik_top}, who examined a double-cone CB and observed a stable equilibrium point gradually shifting towards the equator. This difference stems from the distribution of mass: in our scenario, mass is concentrated at the poles and equator, with lower mass at mid-latitudes. This stands in contrast to the double-cone shape, where mass accumulates primarily at the equator. The presence of an unstable equilibrium point near the pole may prompt the migration of regolith from higher latitudes towards the poles, potentially explaining the observed accumulation of mass at the poles on asteroids like Bennu.

Figure \ref{fig:shape1}b displays the same for a higher friction angle of $20^\circ$ and a shorter initial period of  hours.

Under high rotation rates ($T = 3.37$ hrs) and friction angles ($\delta = 20^\circ$), the equatorial bulge becomes extremely localized, resembling a disc (Fig. \ref{fig:shape1}b). This disc formation echoes recent SPH simulations (\cite{sugiura2021sph}, \cite{hyodo2022formation}) that explored deformation associated with rapid spin-up. The disc shape amplifies tangential gravity while reducing its normal component. This leads to significant mass movement and a substantial decrease in spin due to landslides (see the post-0.8 Myr spin evolution in Fig. \ref{fig:shape1}b).

These results support the findings of \cite{walsh2012spin, sanchezdem, holsapple2010yorp}, which show that beyond a critical rotation rate, further YORP-induced spin increases become impossible. The shape change increases the moment of inertia, slowing the asteroid's spin. While this initially stabilizes slopes, the altered gravity (increased tangential, decreased normal) ultimately destabilizes them. This demonstrates the complex dual effect of landslides on asteroid surface stability.

Importantly, as the disc-like structure grows, it significantly deviates from our initial spherical body assumption, making our theory less applicable. To continue the simulation, we'd need to adopt an alternative reference shape. Furthermore, with the drastically reduced normal gravity in the equatorial bulge, mass shedding becomes prominent.  At this stage, internal failure may also occur, suggesting that a DEM/SPH approach would be more suitable to further study shape evolution.

\begin{figure}[!ht]
    \centering
    \includegraphics[width=\textwidth]{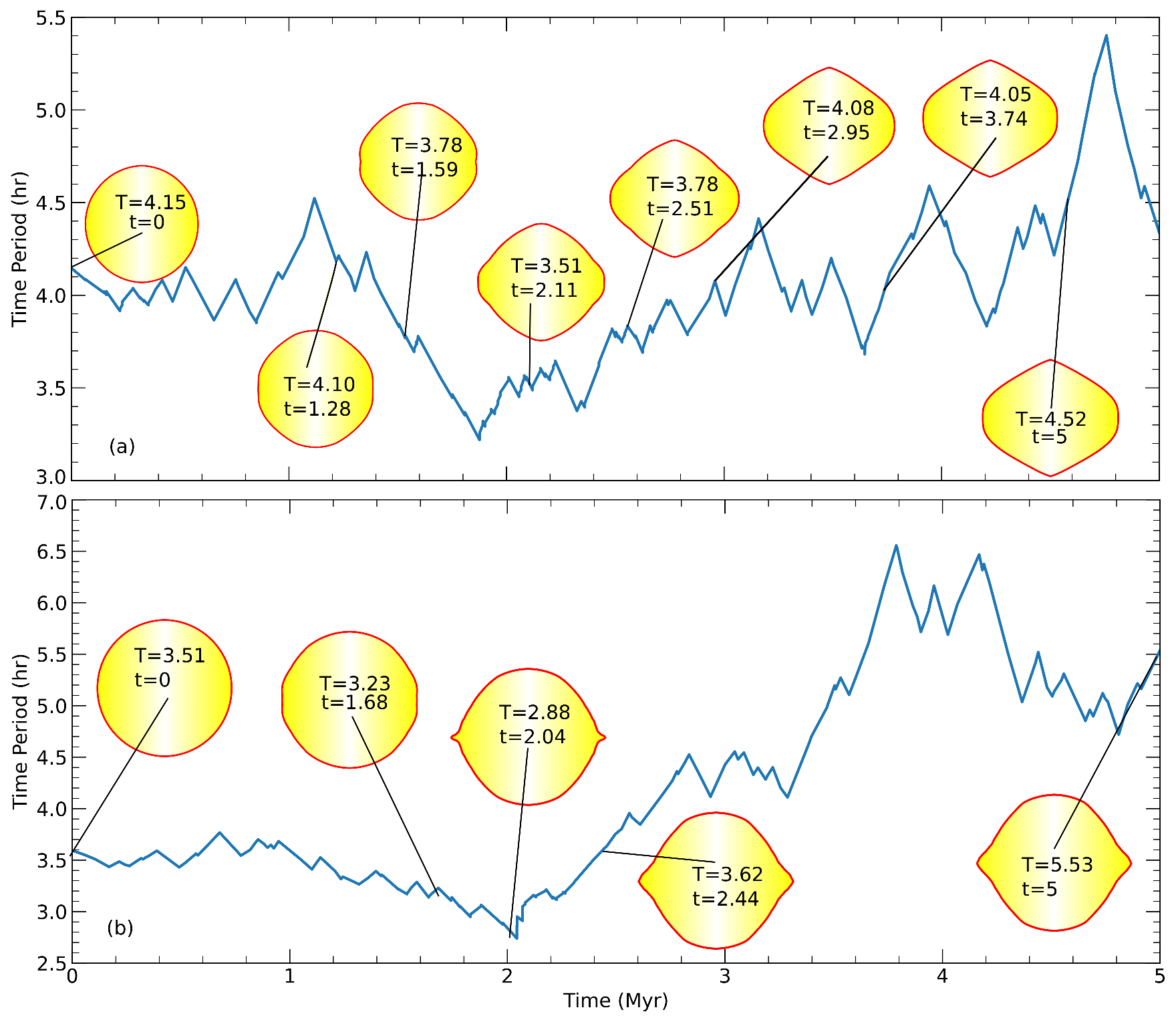}
    \caption{Simulation similar to \ref{fig:shape1} but with stochastic solar torque. (a) Simulation with friction angle $15^\circ$ (b) Simulation with friction angle $30^\circ$. }
    \label{fig:shape2}
\end{figure}

Figure \ref{fig:shape2} portrays a scenario akin to Fig. \ref{fig:shape1}, with a stochastic behaviour of spin change due to solar torque. The emergence of the top shape is delayed, attributed to the comparatively less efficient spinning up process. The stochastic nature of the simulation yields varied outcomes, with instances where top shapes emerge after timescales surpassing the asteroid's lifespan. Figure \ref{fig:shape2}a illustrates the simulation for a friction angle of $15^\circ$, while Fig. \ref{fig:shape2}b corresponds to a friction angle of $30^\circ$. A heightened friction angle consistently yields exaggerated ridge formation or disc-like structures, which may detach from the surface at increased rotation rates, as detailed in \cite{hyodo2022formation}.

\section{Conclusions}
We have derived a set of governing equations to analyze regolith motion on nearly spherical asteroids. This regolith dynamics model is coupled with the asteroid’s angular momentum balance to account for changes in spin rate. The axisymmetric form of the equations is then implemented within the framework presented by \cite{gaurav2025regolith} to investigate the shape and spin evolution of the asteroid in several representative scenarios. Our results suggest that spherical asteroids progressively evolve into top-shaped configurations, with the timescale of this transformation dependent on the nature of the solar torque. A consistently increasing torque accelerates the formation of the top-shape, whereas a stochastic torque leads to a more gradual transition. This shape evolution also gives rise to multiple gravitational equilibrium points on the asteroid’s surface. The proposed theoretical framework can be further extended to investigate regolith dynamics on asteroids with more realistic, irregular geometries.
\bibliographystyle{abbrv} 

\bibliography{bibliography} 
\appendix
\section{Well-posedness of the depth averaged equations}\label{sec:stability}

The system of equations is well posed if the growth rate is bounded for small wavelength perturbations \cite{joseph1990short}. To analyze the well-posedness, we  perturb the base state $(h^0,\,u^0_\theta,\,u^0_\phi)$ and seek solution of the form
\begin{equation}
\begin{pmatrix}
h \\ u_\theta \\ u_\phi
\end{pmatrix}
= e^{i (k\theta-\sigma t)} 
\begin{pmatrix}
\tilde{h} \\ \tilde{u}_\theta \\ \tilde{u}_\phi
\end{pmatrix},
\end{equation}
where $k \gg 1$ is real. Substituting this perturbed field in the governing equations \eqref{eq:cont_fin}-\eqref{eq:vwmom2_fin}, we get
\begin{align}
    \boldsymbol A\cdot \boldsymbol u
    + O\left(\frac{1}{k}\right) =0.
\end{align}
where
\begin{equation}
     \boldsymbol A=
        \begin{bmatrix}
         u_0 -\sigma/k & h_0 & 0 \\
        (\varepsilon \psi h_0 + u_0^2) -\sigma u_0/k & 2 u_0 h_0 -\sigma h_0/k & 0 \\
        u_0v_0-\sigma v_0/k &  v_0 h_0  &  u_0 h_0 -\sigma h_0/k
    \end{bmatrix} \ \ \ \text{and} \ \ \  \boldsymbol u = \begin{pmatrix}
        \tilde{h} \\ \tilde{u} \\ \tilde{v}
    \end{pmatrix}
\end{equation}
For non-trivial solution, the determinant of $\boldsymbol A$ should go to zero as $k\to \infty$. Equating the determinant to zero, gives the characteristic equation of third order whose roots are all real,
\begin{flalign}
    \frac{\sigma}{k} = u_0, u_0 \pm \sqrt{\varepsilon \psi h}
\end{flalign}
and equal to the characteristic speed of the system of hyperbolic PDE \eqref{eq:cont_fin}-\eqref{eq:vwmom2_fin}. Hence upto the order $k$, any perturbation will vary harmonically in time without decay or growth. This is expected, since at the order $k$ ( also known as the principal part) the equation is similar to inviscid fluid equations. The source term which contributes to decay or growth does not appear in the leading order equation. To analyze the growth rate behaviour, we include $O(1)$ terms 
\begin{equation}
    (\boldsymbol A -\frac{i}{k}\boldsymbol a)\cdot \boldsymbol u = 0
\end{equation}
where $\boldsymbol a$ is the contribution from the spatial variation of the base state and due to presence of source terms. The  modified characteristic equation obtained by making det($\boldsymbol A -{i}\boldsymbol a/k$) = 0 is
\begin{equation}
    \left(a_0 + i\frac{a}{k}\right) x^3 + \left(b_0 + i\frac{b}{k}\right) x^2 + \left(c_0 + i\frac{c}{k}\right) x + \left(d_0 + i\frac{d}{k}\right) = 0, 
\end{equation}
where neglecting terms of $o(k)$ is the characteristic equation for det($\boldsymbol A$) = 0 and $x =\omega/k$. The solution of the above cubic polynomial can be expressed as $x=x_0 + \delta x$, where $x_0$ is the solution for the det($\boldsymbol A$) = 0. Linearizing the above equation in $\delta x$, we get
\begin{equation}
    \delta x= \frac{\delta \omega}{k} = \frac{i}{k}\frac{a x_0^3 + b x_0^2 + c x_0 + d}{3a_0 x_0^2 + 2b_0 x_0 + c_0} +O\left(\frac{1}{k^2}\right) \implies {\delta \omega} = {i}\frac{a x_0^3 + b x_0^2 + c x_0 + d}{3a_0 x_0^2 + 2b_0 x_0 + c_0}. 
\end{equation}
Hence the growth/decay rate is independent of $k$ and hence bounded as $k\to \infty$. The system may be linearly stable or unstable depending on the base state and source terms, however, it is not ill-posed  or Hadamard unstable.

\end{document}